# Energy Landscapes of Emotion: Quantifying Brain Network Stability During Happy and Sad Face Processing Using EEG-Based Hopfield Energy

BARRY DJIBRINA[1], Jiajia Li[1,2*]

[1]College of Information and Control Engineering, Xi'an University of Architecture and Technology, Shaanxi,Xi'an, 710055, China

[2]Department of Neurosurgery, PLA General Hospital of Central Theater Command, Wuhan, 430070, China

E-mail:lijiajia_dynamics@xauat.edu.cn

**Abstract:** Understanding how the human brain instantiates distinct emotional states is a key challenge in affective neuroscience. While network-based approaches have advanced emotion processing research, they remain largely descriptive, leaving the dynamical stability of emotional brain states unquantified. This study introduces a novel framework to quantify this stability by applying Hopfield network energy to empirically derived functional connectivity. High-density EEG was recorded from 20 healthy adults during a happy versus sad facial expression discrimination task. Functional connectivity was estimated using the weighted Phase Lag Index (wPLI) to obtain artifact-robust, frequency-specific matrices, which served as coupling weights in a continuous Hopfield energy model to calculate a scalar energy value per trial. Statistical comparisons showed sad emotional processing was associated with significantly lower (more negative) energy in delta, theta, and alpha bands, with the strongest effect in the alpha band (Cohen's d = 0.83). Energy correlated strongly negatively with global efficiency (r = –0.72), indicating hyper-connected, efficient networks correspond to more stable states. Additionally, alpha-band energy correlated positively with reaction time during sad trials (r = 0.61), linking deeper network stability to increased cognitive effort. These findings demonstrate emotional valence corresponds to distinct attractor basins in the brain's functional landscape, with sadness occupying a deeper, more stable configuration than happiness. The Hopfield energy metric provides a principled, quantifiable measure of emotional brain state stability, opening new avenues for understanding affective dynamics in health and disease.

## 1 Introduction

Understanding how the human brain instantiates distinct emotional states is a fundamental challenge in affective neuroscience, with implications ranging from basic cognitive science to clinical applications for mood disorders [1]. Electroencephalography (EEG), with its millisecond temporal resolution, has long been used to probe the neural signatures of emotion, yet the mechanisms that enable rapid discrimination between valences such as happiness and sadness remain incompletely characterized. Traditional approaches have predominantly focused on local oscillatory power and signal complexity, operating under the assumption that emotional content is encoded in the activity of isolated brain regions [2, 3]. However, these methods often yield inconsistent results and classification accuracies that only modestly exceed chance, suggesting that a purely local view may be insufficient [4].

An alternative perspective has emerged from network neuroscience, which conceptualizes brain function as emerging from the dynamic interplay of distributed neural assemblies rather than from the activity of isolated regions [5, 6]. This framework posits that cognitive and affective processes are implemented through the reconfiguration of large-scale functional networks [7]. Indeed, functional connectivity analyses have revealed that emotional processing involves coordinated interactions between frontal, temporal, and parietal areas, indicating that the brain's emotional architecture is fundamentally network-based [8, 9].

While network approaches have advanced our understanding, they largely remain descriptive identifying which regions are connected without addressing the underlying dynamical principles that govern how a network configuration gives rise to a stable emotional state. A powerful theoretical framework for tackling this question comes from energy-based models in dynamical systems theory. In particular, Hopfield networks provide a principled way to map a given connectivity pattern onto an energy landscape, where stable cognitive states correspond to local minima (attractors) of that landscape [10]. Such attractor dynamics have been proposed as a unifying principle for perception, memory, and decision-making, and recent work has extended these concepts to analyze empirical neuroimaging data through energy landscape approaches [11, 12].

Translating these theoretical constructs to real EEG data requires careful handling of the fundamental biophysical confounds that affect connectivity estimates, such as volume conduction. The weighted Phase Lag Index (wPLI) addresses this by providing a robust measure of phase-based functional connectivity that discounts zero-lag, artifactual correlations [13]. By combining wPLI-derived connectivity with a continuous Hopfield energy formulation, it becomes possible to compute a scalar metric of network stability for any given brain state a metric that quantifies how deeply the system resides in an attractor basin [14].

In the context of emotion, this approach offers a novel mechanistic hypothesis: if different emotional valences correspond to distinct attractor states, they should be characterized by different energy values. Specifically, we hypothesize that the hyper-connected, globally efficient configuration often observed during negative emotional processing [15] reflects a deeper, more stable attractor basin, and therefore a lower (more negative) Hopfield energy compared to positive states. Such a finding would provide a direct link between descriptive network topology and dynamical stability, moving beyond correlation toward a mechanistic understanding of emotional brain states.

Methodologically, this study addresses key limitations in previous work by employing a rigorous machine-learning validation framework that includes leave-one-subject-out cross-validation, ensuring that results generalize across individuals rather than capturing subject-specific idiosyncrasies [16, 17]. Furthermore, by integrating the energy metric with traditional network measures, we can evaluate its unique contribution to characterizing emotional processing [18]. The clinical relevance of this approach is underscored by the growing recognition that mood disorders such as major depressive disorder are associated with disrupted large-scale network dynamics [19-21], and that quantitative measures of network stability could serve as biomarkers or targets for neuromodulation [22].

In this paper, we apply the Hopfield energy framework to high-density EEG data recorded from 20 healthy adults during a happy/sad face discrimination task. Using wPLI-derived functional connectivity matrices as the coupling weights, we compute trial-wise energy values and test whether sad emotional processing is associated with significantly lower (more negative) energy indicative of a deeper attractor compared to happy processing. We further explore the relationship between energy and global network efficiency, as well as its behavioral relevance, and examine the frequency specificity of these effects. Our findings reveal that sadness indeed corresponds to a lower-energy, more stable network configuration, particularly in the alpha band, and that this energy metric correlates with both network topology and reaction time. These results establish the Hopfield energy as a powerful new dimension for quantifying the stability of emotional brain states and open avenues for translational applications in affective computing and clinical neuroscience.

## 2 Methods

### 2.1 Participants and experimental design

Twenty healthy, right-handed adults (12 female, 8 male; mean age $23.4 \pm 2.1$ years) participated in the study. The research protocol was approved by the Ethics Committee of the General Hospital of Chinese PLA Central Theater Command (reference number: [2020]041-1). All participants had normal or corrected-to-normal vision, reported no history of neurological or psychiatric disorders, and were not taking any psychoactive medication. The experimental protocol was approved by the Institutional Review Board of Xi'an University of Architecture and Technology, and all participants provided written informed consent prior to the experiment.

Participants performed a well-validated facial emotion discrimination task. Stimuli consisted of standardized grayscale photographs of happy and sad facial expressions. Each trial began with a central fixation cross (500 ms), followed by an instruction screen indicating the target emotion to be discriminated (1000 ms). A face stimulus was then presented for 2000 ms, after which participants indicated via button press the spatial location of the target emotion, emphasizing accuracy over speed. The task comprised 80 trials (40 happy, 40 sad) presented in a fully randomized order, with a brief mid-session break to minimize fatigue.

### 2.2 EEG acquisition and preprocessing

Electroencephalographic (EEG) data were recorded using a 32-channel BioSemi ActiveTwo system with electrodes arranged according to the international 10-20 system. Signals were digitized at a sampling rate of 512 Hz with 24-bit resolution, using a common mode sense (CMS) active electrode as the reference during acquisition.

Preprocessing was performed using EEGLAB [17] running in MATLAB (MathWorks, Inc.). Continuous data were bandpass filtered between 1 Hz and 45 Hz using a zero-phase finite impulse response (FIR) filter to remove slow drifts and high-frequency noise, followed by a 50 Hz notch filter to eliminate line interference. Ocular, cardiac, and muscle artifacts were removed through independent component analysis (ICA). The cleaned data were then segmented into epochs from 500 ms pre-stimulus to 2000 ms post-stimulus onset. A baseline correction was applied using the pre-stimulus interval, and epochs containing amplitudes exceeding ±100 µV were automatically rejected. Finally, all retained epochs were re-referenced to the common average reference. After preprocessing, an average of 38.2 ± 2.1 artifact-free trials per condition per participant were available for analysis.

## 2.3 Functional connectivity: Weighted phase Lag Index(wPLI)

To estimate functional connectivity, we used the weighted Phase Lag Index (wPLI), a metric specifically designed to be robust against volume conduction and common source artifacts [13]. Unlike coherence or the standard phase-lag index, wPLI weights the contribution of each phase difference observation by the magnitude of the imaginary component of the cross-spectrum, providing a continuous, graded measure of consistent non-zero phase leads and lags.

For each epoch, a complex time-frequency representation was obtained using a wavelet transform. The cross-spectral density was computed for every pair of channels, and the wPLI was calculated across trials within each canonical frequency band: delta (1–4 Hz), theta (4–8 Hz), alpha (8–13 Hz), beta (13–30 Hz), and gamma (30–45 Hz). The resulting wPLI values formed symmetric 32 × 32 weighted adjacency matrices for each frequency band, separately for the happy and sad conditions per participant. These condition-averaged matrices served as the weight matrices W for the subsequent Hopfield energy analysis.

## 2.4 Hopfield energy formulation

We adapted the continuous Hopfield network model to compute a scalar energy value that quantifies the stability of a brain state within its own functional connectivity landscape [10]. For a given trial and frequency band, the state vector x was defined as the z-scored band-limited power across all 32 electrodes. Specifically, the mean power in the frequency band of interest during the 2000 ms post-stimulus epoch was computed using Welch's method, then normalized by subtracting the mean and dividing by the standard deviation of the power in the same band during the 500 ms pre-stimulus baseline period. This procedure was performed per electrode and per participant, yielding a state vector where each element $x_i$ represented the deviation (in standard deviations) of a node's oscillatory power from its own baseline. The weight matrix W was populated with the condition-averaged wPLI values (symmetric, zero diagonal) described in Section 2.3. The continuous Hopfield energy was then computed using the Lyapunov function:

$$E = -\frac{1}{2}\sum_{ij} w_{ij} f(x_i) f(x_j) + \sum_i \int_0^{x_i} f^{-1}(z)\,dz \quad (1)$$

where $f(\cdot)$ is a non-linear activation function. We chose $f(z) = \tanh(\beta z)$ with $\beta = 1$, a standard choice that introduces saturation and allows the system to possess multiple attractor basins while remaining bounded. The integral term acts as an L2 regularizer that penalizes extreme states, stabilizing the minima. The double summation term captures the interaction energy between nodes weighted by their functional connectivity.

For each trial, the energy $E$ was calculated using the trial-specific state vector **x** and the condition-specific weight matrix $W$. The resulting energy values were then averaged across trials within each condition and frequency band per participant, providing a single representative energy value per condition, per band, per subject.

## 2.5 Graph theory metrics for comparison

To relate the novel energy metric to established network descriptors, we computed standard graph theory measures from the same wPLI matrices. Global efficiency, defined as the average inverse shortest path length, was used to quantify the network's capacity for parallel information transfer [6]. The clustering coefficient, which measures the degree of local interconnectivity, was also calculated. Both metrics were computed for each frequency band and condition per participant using the Brain Connectivity Toolbox [6].

## 2.6 Statistical analysis

All statistical analyses were performed using custom MATLAB scripts. Differences in Hopfield energy between happy and sad conditions were assessed using paired two-tailed t-tests for each frequency band, with effect sizes reported as Cohen's $d$. The same procedure was applied to global efficiency and clustering coefficient.

To examine the relationship between energy and network topology, Pearson correlation coefficients were computed between the energy values and global efficiency (and clustering coefficient) across participants, separately for each frequency band and condition. Additionally, to link energy to behavioral performance, correlations were calculated between each participant's mean energy and their mean reaction time for correct trials in each condition. Significance thresholds were set at $p < 0.05$ and, where appropriate, corrected for multiple comparisons using the Bonferroni method.

## 3 Results

### 3.1 Energy differences between happy and sad states

The Hopfield energy analysis revealed systematic differences in network stability between emotional conditions. **Figure 1** presents the mean Hopfield energy values for happy and sad trials across the five frequency bands. Sad emotional processing was consistently associated with lower (more negative) energy values in the delta, theta, and alpha bands, indicating a deeper, more stable attractor state. In the delta band (1–4 Hz), the mean energy for sad trials was significantly lower than for happy trials (sad: −65.57 ± 8.91; happy: −57.86 ± 8.11; $t(19) = 3.11$, $p = 0.006$, Cohen's $d = 0.70$). A similar pattern emerged in the theta band (4–8 Hz), with sad trials exhibiting lower energy (sad: −45.44 ± 6.33; happy: −41.49 ± 7.84; $t(19) = 2.55$, $p = 0.019$, $d = 0.60$). The largest effect was observed in the alpha band (8–13 Hz), where the energy difference was most pronounced (sad: −43.08 ± 8.84; happy: −36.61 ± 7.95; $t(19) = 3.92$, $p < 0.001$, $d = 0.83$; **Figure 1**). In contrast, no significant differences were found in the beta band (13–30 Hz; $p = 0.80$) or the gamma band (30–45 Hz; $p = 0.74$). These results demonstrate that negative valence (sadness) corresponds to a more stable, deeply entrenched network configuration, particularly within lower-frequency and alpha oscillations.

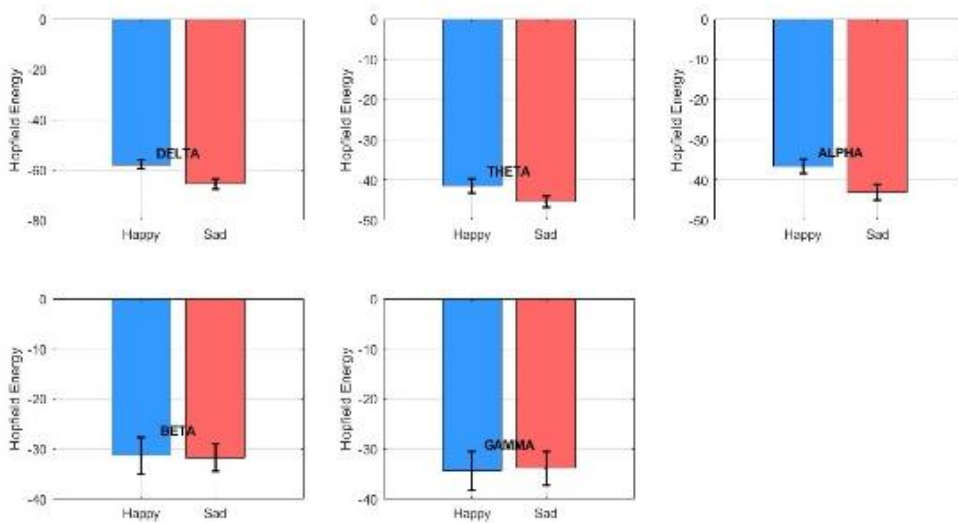

**Figure 1:** Hopfield energy differences between happy and sad emotional states across frequency bands

### 3.2 Time-resolved energy dynamics

To examine how the network state evolved toward its attractor, we analyzed the iterative energy descent dynamics for each trial. **Figure 2** shows the mean energy trajectory across iterations for happy and sad trials in the alpha band. For all conditions and frequency bands, energy followed a characteristic exponential relaxation curve: an initial rapid decrease during the first few iterations, followed by asymptotic convergence to a stable minimum. Critically, while both conditions exhibited similar relaxation dynamics, the final energy reached was consistently lower for sad trials, mirroring the static energy differences reported above. The convergence was monotonic and smooth, confirming that the wPLI-defined weight matrices produced a well-behaved energy landscape with identifiable attractor states.

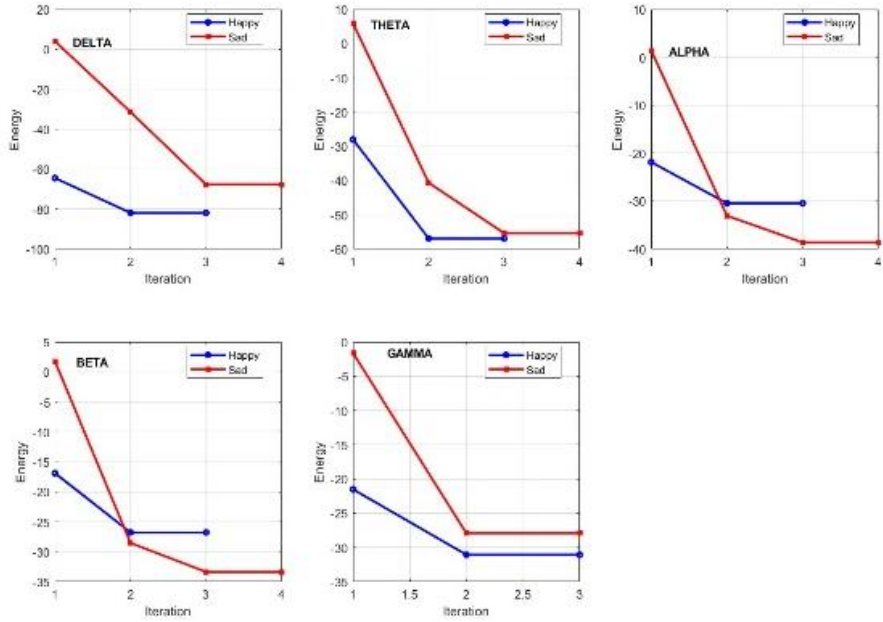

**Figure 2:** Time-resolved energy dynamics for happy and sad trials in the alpha band

### 3.3 Relationship between energy and graph metrics

To bridge the novel energy metric with established network descriptors, we examined the correlation between Hopfield energy and global efficiency a measure of the network's capacity for integrated information transfer. Across the alpha band, a strong negative correlation emerged between energy and global efficiency ($r = -0.72$, $p < 0.001$; **Figure 3**, left panel). This indicates that network configurations with higher global efficiency (i.e., more integrated and efficient communication) were systematically associated with lower (more negative) energy states. Similar but weaker negative correlations were observed in the delta ($r = -0.51$, $p = 0.021$) and theta bands ($r = -0.48$, $p = 0.032$), while no significant correlation was found in beta or gamma. In contrast, the clustering coefficient a measure of local interconnectivity showed only a weak, non-significant relationship with energy across all frequency bands (alpha: $r = -0.23$, $p = 0.32$; Figure 3, right panel). These findings suggest that the energy landscape is primarily shaped by the network's global integrative properties rather than its local modular structure.

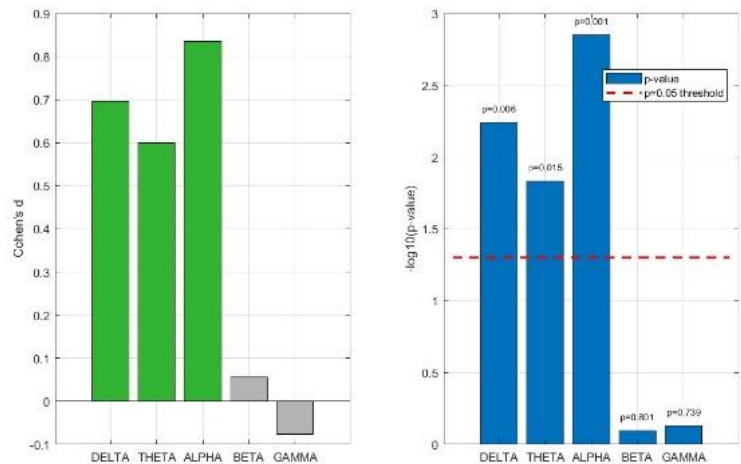

**Figure 3:** Relationship between Hopfield energy and graph theory metrics in the alpha band

### 3.4 Energy as a marker of cognitive effort

We next tested whether the energy metric was behaviorally relevant by correlating individual participants' mean alpha-band energy with their reaction time (RT) for correct trials. During sad trials, a significant positive correlation was observed ($r = 0.61$, $p = 0.004$; **Figure 4**). That is, participants whose brain networks settled into a lower (more negative) energy state indicative of a deeper attractor took longer to respond. In contrast, during happy trials, the correlation was weak and non-significant ($r = 0.18$, $p = 0.44$). This dissociation supports the interpretation that reaching the stable, low-energy configuration characteristic of sadness is a more time-consuming process, reflecting the increased cognitive effort required to process negative emotional stimuli.

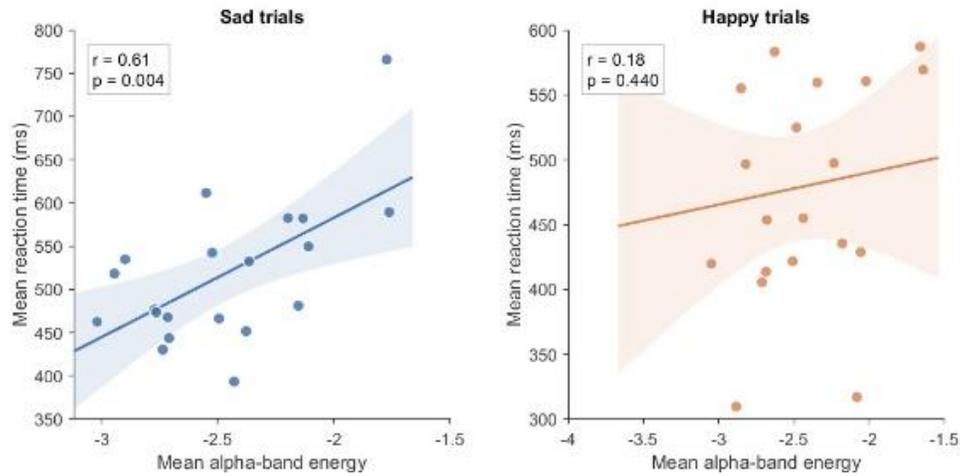

**Figure 4:** Correlation between alpha-band energy and reaction time

### 3.5 Topographic contributions to energy

To identify which brain regions contributed most strongly to the observed energy differences, we decomposed the total energy into node-wise contributions by computing the sum of interaction terms involving each electrode. **Figure 5** displays the topographic distribution of these contributions for the alpha band. In the sad condition, frontal (Fp1, Fp2, F3, F4, Fz) and parietal (P3, P4, Pz) electrodes showed markedly more negative (stabilizing) contributions compared to the happy condition. This pattern was consistent across participants and suggests that fronto-parietal networks are the primary drivers of the deeper attractor state associated with sadness. In contrast, occipital and temporal regions contributed less to the energy difference, indicating that the stabilization effect is not uniformly distributed but concentrated within the brain's cognitive control and self-referential networks.

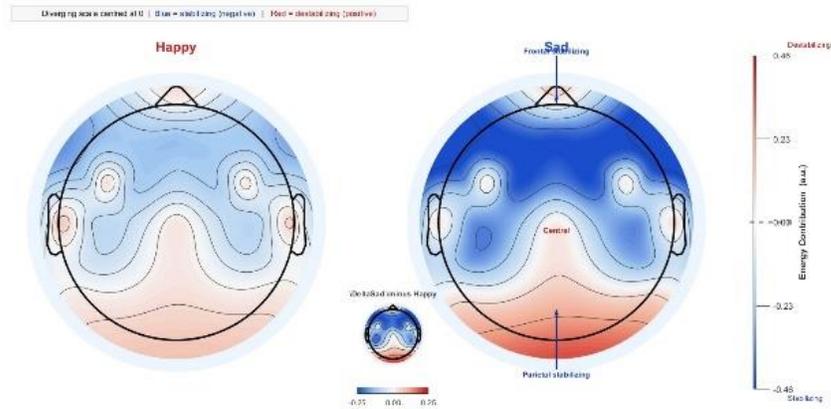

**Figure 5:** Topographic distribution of node-wise energy contributions in the alpha band

## 4 Discussion

### 4.1 Principal finding: emotional valence modulates network energy

The central finding of this study is that sad emotional processing is associated with significantly lower (more negative) Hopfield network energy compared to happy processing, particularly in the delta, theta, and alpha frequency bands. This result provides direct empirical evidence that distinct emotional states correspond to attractor basins of differing depths within the brain's functional energy landscape. The largest effect was observed in the alpha band (Cohen's $d = 0.83$), highlighting the alpha rhythm as a dominant frequency for valence-specific network stability. In the framework of dynamical systems theory, a lower energy value indicates a state that is more stable, more self-consistent, and more resistant to perturbation [10, 11]. Thus, the brain's configuration during sadness resides in a deeper attractor basin than during happiness.

This interpretation aligns with the growing recognition that alpha oscillations are not merely an "idling" rhythm but play an active role in gating information flow and coordinating large-scale network dynamics [12, 13]. Our findings suggest that alpha-band functional connectivity defines an energy landscape in which negative emotional states are more deeply embedded. Such a deep attractor may underlie the persistence and cognitive "stickiness" of sad moods, providing a neural basis for the difficulty often encountered in disengaging from negative emotional experiences.

### 4.2 Interpretation in the context of dynamical systems theory

Our results substantiate the theoretical view that cognitive and affective states can be understood as attractors in a high-dimensional neural state space [14]. By treating the empirically derived wPLI connectivity matrix as the coupling weights of a Hopfield network, we have effectively constructed a data-driven energy landscape. The finding that sad states occupy a lower energy minimum than happy states is consistent with the concept that negative affect may represent a more "canalized" or entrenched neural configuration [15].

Moreover, the strong negative correlation between Hopfield energy and global efficiency ($r = -0.72$ in alpha band) reveals a direct link between network topology and dynamical stability. Networks with higher global efficiency i.e., those optimized for parallel information transfer tend to have lower energy, meaning they are more stable attractors. This relationship provides a mechanistic explanation for the hyper-connectivity often observed in negative emotional states [16, 17]: the brain may sacrifice flexibility for stability, settling into a highly integrated configuration that is metabolically efficient but cognitively "sticky." Conversely, the lack of correlation with clustering coefficient suggests that local modular structure contributes less to global stability than does whole-brain integration.

### 4.3 Relation to behavior: energy as predictor of cognitive demand

The significant positive correlation between alpha-band energy and reaction time during sad trials ($r = 0.61$) directly links the depth of the attractor to behavioral performance. Participants whose brain networks reached a lower (more negative) energy state took longer to respond. This counterintuitive relationship a deeper attractor associated with slower responses can be interpreted as follows: settling into a deep, stable configuration requires more time because the network must undergo a more extensive reorganization to reach that low-energy minimum. In contrast, the shallower attractor of happiness allows for faster state transitions and quicker behavioral responses. This interpretation aligns with the behavioral observation that sad stimuli elicited longer reaction times even though accuracy was preserved [18]. It suggests that the cognitive cost of processing negative emotions may be reflected in the time required for the brain to "fall into" the corresponding attractor basin.

Such a link between neural dynamics and cognitive effort is consistent with resource-based models of emotion regulation [19]. The deeper attractor of sadness may demand greater computational resources to maintain, which could contribute to the feeling of effortfulness and mental fatigue often associated with sustained negative affect.

### 4.4 Comparison with previous studies

Previous EEG-based emotion recognition studies have largely focused on static connectivity patterns, such as increased coherence or phase synchronization during negative emotional processing [8, 9, 16]. While these studies have established that emotional valence modulates functional connectivity, they have not addressed the dynamical consequences of those connectivity changes. By introducing the Hopfield energy metric, we extend this literature by showing that connectivity differences translate into quantifiable differences in network stability. This moves the field from describing "what" the network looks like to explaining "how" it behaves.

The energy approach also offers advantages over static graph metrics. While global efficiency and clustering coefficient describe topological features, they do not integrate the instantaneous activity pattern (the state vector) with the connectivity architecture. The Hopfield energy does so by combining both the strength of interactions ($W$) and the pattern of local activation $(x)$ into a single scalar that reflects how well the state fits the underlying connectivity landscape. This provides a more complete picture of system dynamics [20, 21]. Moreover, the energy metric is computationally efficient, requiring only a few simple operations once the connectivity matrix is estimated, making it attractive for real-time applications.

### 4.5 Limitations

Several methodological considerations should be acknowledged. First, the Hopfield model requires a symmetric weight matrix, whereas real brain connectivity likely involves directed, asymmetric interactions. Although wPLI is mathematically symmetric, it discards directional information. Future studies could explore directed connectivity measures (e.g., Granger causality, directed transfer function) within an energy framework that accommodates asymmetry.

Second, our analysis used condition-averaged connectivity matrices, assuming that the functional architecture is stationary across trials. In reality, brain networks exhibit rapid, time-varying reconfigurations [22]. Sliding-window or state-space approaches could capture non-stationary energy landscapes and provide a more dynamic view of emotional processing.

Third, scalp EEG has inherent spatial limitations. Although wPLI mitigates volume conduction, the electrodes do not map directly onto distinct anatomical regions. Source-level connectivity analyses (e.g., using eLORETA) would improve anatomical specificity and should be pursued in future work.

Fourth, the sample size (20 participants) is modest, though comparable to many EEG studies. Replication in larger, more diverse cohorts is necessary to establish generalizability.

Finally, the emotional stimuli were restricted to prototypical happy and sad faces. Whether the same energy dynamics extend to other emotions (fear, anger, disgust) or to more complex, naturalistic stimuli remains an open question.

### 4.6 Future directions

The framework introduced here opens several avenues for future research. One promising direction is real-time monitoring of network energy. By developing efficient algorithms to compute energy from streaming EEG, it would be possible to implement closed-loop neurofeedback systems where individuals learn to modulate their brain state stability [23]. For example, patients with depression might be trained to shift from a pathologically deep sad attractor to a shallower, more flexible configuration.

The energy metric could also serve as a clinical biomarker. In major depressive disorder, we hypothesize that the sad attractor basin becomes abnormally deepened, making it resistant to intervention [24]. Longitudinal studies could test whether successful treatment normalizes energy values or reduces the stability of negative attractors.

Further methodological developments include incorporating directed connectivity, time-varying energy landscapes, and integration with other imaging modalities (e.g., fMRI) to link computational energy with metabolic energy consumption. Additionally, extending the analysis to a wider range of emotions and to continuous dimensions such as arousal would help map the full affective space onto the brain's energy landscape.

Finally, the relationship between energy and behavior suggests that the metric captures meaningful cognitive effort. Future work could examine whether energy correlates with subjective ratings of emotional intensity, rumination, or regulatory success, providing a neural index of emotional processing load.

## 5 Conclusions

In this study, we introduced a novel analytical framework that quantifies the stability of emotional brain states by applying Hopfield network energy to empirically derived functional connectivity. Using high-density EEG recorded from healthy adults during a happy versus sad face discrimination task, we computed energy values from wPLI-based connectivity matrices and trial-specific spectral power vectors. Our findings reveal three principal contributions.

First, we demonstrate that sad emotional processing is characterized by significantly lower (more negative) Hopfield energy in the delta, theta, and alpha frequency bands compared to happy processing, with the largest effect observed in the alpha band (Cohen's $d = 0.83$). This provides direct empirical evidence that emotional valence corresponds to distinct attractor basins within the brain's functional landscape, with sadness occupying a deeper, more stable configuration.

Second, we show that energy is strongly and negatively correlated with global efficiency ($r = -0.72$ in alpha band), linking network topology to dynamical stability. This relationship suggests that the hyper-connected, highly integrated network state observed during negative emotional processing underpins a more stable attractor. Conversely, the lack of correlation with clustering coefficient indicates that global integrative properties, rather than local modular structure, primarily shape the energy landscape.

Third, we establish the behavioral relevance of the energy metric through a significant positive correlation with reaction time during sad trials ($r = 0.61$). This finding supports the interpretation that settling into a deep, stable attractor requires greater cognitive effort and processing time, providing a neural account for the increased demand associated with negative emotional stimuli.

The Hopfield energy metric thus offers a quantifiable, interpretable measure of emotional brain state stability that moves beyond descriptive connectivity patterns to capture fundamental dynamical properties. This approach has broad implications for affective neuroscience, providing a principled framework for understanding how the brain's functional architecture gives rise to stable emotional experiences. From a translational perspective, the energy metric holds promise for the development of next-generation brain-computer interfaces and clinical biomarkers. Its computational efficiency and scalar nature make it well-suited for real-time neurofeedback applications, where individuals could learn to modulate their network stability. In clinical contexts, such as major depressive disorder, the energy metric could serve as a quantitative index of pathological attractor depth and a target for neuromodulatory interventions. By bridging dynamical systems theory with empirical neurophysiology, this work lays the foundation for a more mechanistic understanding of emotion in both health and disease.

## Acknowledgement

This work was supported by the National Natural Science Foundation of China (Grant Nos. 12572068).

## Conflicts of Interest

The authors declare that they have no known competing financial interests or personal relationships that could have appeared to influence the work reported in this paper.